\documentclass[journal]{IEEEtran}
\IEEEoverridecommandlockouts 

\usepackage{cite}

\ifCLASSINFOpdf
   \usepackage[pdftex]{graphicx}
   \graphicspath{{./Images/}}
   \DeclareGraphicsExtensions{.pdf,.jpeg,.png}
\else
\fi
\usepackage[cmex10]{amsmath}
\makeatother

\usepackage{hyphenat}
\usepackage[cmex10]{amsmath}
\usepackage{amssymb}
\usepackage{verbatim}
\usepackage{amsfonts}
\usepackage{subfig}
\usepackage[update,prepend]{epstopdf}
\usepackage{float}
\usepackage[usenames,table,dvipsnames]{xcolor}
\usepackage[hidelinks]{hyperref}
\usepackage{tabularx}
\usepackage{bm}
\usepackage{siunitx}
\usepackage{enumerate}
\usepackage{multirow}
\usepackage{booktabs,colortbl}
\usepackage{algorithm,algpseudocode}
\usepackage{setspace}
\usepackage{diagbox}
\usepackage{multirow}
\usepackage{mathrsfs}

\allowdisplaybreaks

\begin{document}
\title{New Cycle-based Formulation, Cost Function, and Heuristics for DC OPF Based Controlled Islanding}
\author{Ilya~Tyuryukanov,
		Marjan~Popov,          
        Jorrit A. Bos,       
        Mart A. M. M.~van~der~Meijden,
        and~Vladimir~Terzija  
\thanks{I. Tyuryukanov and M. Popov are with Delft University of Technology, Delft 2628CD, Netherlands. Since October 1, 2021, I. Tyuryukanov is also with Siemens Energy AG, Freyeslebenstr. 1, Erlangen 91058, Germany (e-mail: ilya.tyuryukanov@ieee.org, m.popov@tudelft.nl).}%
\thanks{J. Bos and M.\,A.\,M.\,M.~van~der~Meijden are with TenneT TSO B.V., Utrechtseweg 310, Arnhem 6812AR, Netherlands (e-mail: \{Jorrit.Bos, mart.vander.meijden\}@tennet.eu). M.\,A.\,M.\,M.~van~der~Meijden is also with Delft University of Technology, Delft 2628CD, Netherlands.}%
\thanks{V. Terzija is with Key Laboratory of Power System Intelligent Dispatch and Control of Ministry of Education, School of Electrical Engineering, Shandong University, Jinan 250061, China (email: terzija@ieee.org)}}%

\markboth{Journal of \LaTeX\ Class Files,~Vol.~14, No.~8, August~2015}%
{Shell \MakeLowercase{\textit{et al.}}: Bare Demo of IEEEtran.cls for IEEE Journals}
\maketitle

\begin{abstract}
This paper presents a new formulation for intentional controlled islanding (ICI) of  power transmission grids based on mixed-integer linear programming (MILP) DC optimal power flow (OPF) model. We highlight several deficiencies of the most well-known formulation for this problem and propose new enhancements for their improvement. In particular, we propose a new alternative optimization objective that may be more suitable for ICI than the minimization of load shedding, a new set of island connectivity constraints, and a new set of constraints for DC OPF with switching, and a new MILP heuristic to find initial feasible solutions for ICI. It is shown that the proposed improvements help to reduce the final optimality gaps as compared to the benchmark model on several test instances. 
\end{abstract}

\section{Introduction}
\label{sec:intro}
Intentional controlled islanding (ICI) is an important system integrity protection scheme (SIPS) aiming to prevent system collapse due to wide-area instability by separating the system into a set of non-interacting islands. A typical situation requiring such a control action is loss of synchronism between some parts of the system, but it can also be used to limit the spread of cascading outages, or to isolate unstable parts of the system. The high relevance of this control action is confirmed by the increasing number of blackouts around the globe \cite{Alhelou.2019}.

From the mathematical perspective, computing the splitting boundary to separate the system into a number of self-sustained islands equates to an OPF problem with switching. Such problems are notoriously hard to solve, as OPF is nonlinear, and switching decisions are discrete, which amounts to a mixed-integer nonlinear program (MINLP). Moreover, ICI must be solved rapidly in real-time as a corrective control action for instability mitigation. For this reason, solving ICI as a MINLP is not feasible, and  MILP reformulations of the original MINLP are often seen as an acceptable trade-off between accuracy and computation time. 

Although multiple MILP approximations for ICI and optimal transmission switching (OTS) exist in literature (e.g., \cite{Trodden.2014,Brown.2020}), this paper is focused on the basic DC OPF ICI model. Our goal is to resolve the fundamental issues associated with the DC OPF ICI model to devise the efficient computational enhancements that could also be applied to more advanced MILP-based ICI models \cite{Trodden.2014,Coffrin.2015,Brown.2020}. 

We first observe that the conventional ICI objective of minimizing the total post-splitting load shedding \cite{Fan.2012,Teymouri.2017} may often result in a poor MIP optimality gap progression if the exact optimal load shedding is close to zero. To alleviate this issue, we propose to minimize the total load-generation imbalance in the computed islands while possibly keeping the load shedding as the secondary optimization objective. Minimizing the total load-generation imbalance should limit load shedding and additionally limit the value of the rate of change of frequency (ROCOF) following the system splitting.

Our second contribution aims to improve the modeling of the islands' connectedness requirement. The previously published papers on optimization-based ICI either use artificial commodity flows to impose connectivity constraints on islands \cite{Teymouri.2017} or neglect this requirement \cite{Trodden.2013,Trodden.2014}. Our proposal is to model connectivity through directed spanning trees, which aims to avoid the computationally inefficient \emph{big-M} coefficients that accompany the commodity flow based formulation. This technique has previously been applied to enforce radiality in distibution networks \cite{Jabr.2013}, but our approach differs from \cite{Jabr.2013}. 

Our third contribution addresses the big-M coefficients that are normally present in the DC OPF constraints related to switching decisions. We reformulate these constraints as  Kirchhoff's voltage law (KVL) equations with additional slack variables to eliminate the uncertain voltage angle differences across an open line otherwise modeled as large big-M constants. Cycle inequalities involving KVL have been previously proposed in \cite{Kocuk.2016}, but our work differs from \cite{Kocuk.2016} in several aspects. The goal of \cite{Kocuk.2016} was to strengthen the LP relaxation of the conventional DC optimal transmission switching (OTS) model, while our goal is the elimination of the big-M coefficients. This motivates a different modeling approach. 

Our fourth contribution is a new MILP heuristic to find initial feasible solutions for DC OPF ICI. For some MIP problems, a feasible solution can be obtained trivially or in polynomial time, but the  OPF ICI problem combining power flow physics and discrete decisions does not allow for a simple initialization. Obtaining an early feasible solution is crucial for the progress of a modern MIP solver, as there are efficient heuristics that are able to progressively improve the best feasible solution (e.g., the RINS heuristic). Unfortunately, we could not identify any published MILP heuristics for OPF ICI, which prompted the design of a simple yet effective method based on the LP relaxation of DC OPF ICI. 

The following sections describe the above contributions in more detail. Section \ref{sec:milp_basic} introduces the used notation and the conventional benchmark DC OPF ICI model. Section \ref{sec:milp_our} details our proposed model and cost function. Section \ref{sec:milp_heur} introduces the MILP heuristic for OPF ICI and explains how MILP is solved. Section \ref{sec:results} describes the test cases and presents the computational results. Section \ref{sec:conclu} concludes the paper.

\section{Overview of Previous Results}
\label{sec:milp_overview}
\subsection{Notations}
\label{sec:notation}
A power network consisting of $n$ nodes and $m$ branches is represented by a graph with set of nodes $\mathcal{V}$, set of edges $\mathcal{E}$ ($|\mathcal{E}|=m$), and set of arcs $\mathcal{A}$ ($|\mathcal{A}|=2m$). Power demand and power generation at node $i$ prior to splitting are denoted as $P_{L,i}^s$ and $P_{G,i}^s$. Load demand and generation shedding to satisfy the post-splitting conditions are denoted as $P_{L,i}$ and $P_{GS,i}$. Power flow through branch $(i,j)\in\mathcal{E}$ is defined as $p_{i,j}$, its value prior to splitting is $p_{i,j}^s$, and its limiting magnitude is $p_{i,j}^{max}$. Admittance of branch $(i,j)$ is denoted as $b_{i,j}$. As the used model is DC OPF, reactive power is not modeled.

The splitting configuration is modeled by binary variables $x_{i,k}$ ($x_{i,k}=1$ implies that node $i$ is assigned to island $k$) and $y_{i,j}$ ($y_{i,j}=1$ implies that branch $(i,j)\in\mathcal{E}$ is open). The total number islands is $K$.  An important ICI requirement that promotes transient stability after splitting is to assign generators from the same coherent group to one island. Coherent groups are denoted as $\mathcal{G}_1,\ldots,\mathcal{G}_K$  (i.e., each coherent group is supposed to form a separate island). 

Indices $i$ and $j$ run over network nodes and pairs of nodes (edges or arcs). Index $k$ runs over islands ($k=1,\ldots,K$). Superscripts $s$, $*$, $LR$ denote pre-splitting values, integral solutions of MILP programs, and linear relaxation (LR) solutions of MILP programs respectively.

\subsection{Basic MILP Model for DC OPF ICI}
\label{sec:milp_basic}
Probably the first paper proposing a MILP-based ICI strategy was \cite{Fan.2012}. Since then several related models appeared in the literature, of which the model in \cite{Trodden.2013} is the most compact one. Unlike \cite{Fan.2012,Kyriacou.2017,Patsakis.2019}, the model in \cite{Trodden.2013} does not use auxiliary binary variables that model the membership of edge $(i,j)$ in island $k$. Instead, the edge status variables $y_{i,j}$ are used directly, which eliminates  $O(mK)$ extra binary variables and constraints. However unlike most other references, the models in \cite{Trodden.2013,Trodden.2014} neglect the islands' connectedness requirement. To avoid the deficiencies of any single existing model, a combined DC OPF ICI model is introduced below:

\begin{subequations}
\begin{align}
&\text{min}\quad \beta\sum_{i\in\mathcal{V}}P_{LS,i} + \gamma\sum_{i\in\mathcal{V}}P_{GS,i} + \mu\sum_{(i,j)\in\mathcal{E}}p_{i,j}^s y_{i,j}\label{eqn:objective1}\\
&\sum_{k=1}^K x_{i,k}=1,~\forall i \in \mathcal{V} \label{eqn:dc_xik}\\
& x_{i,k}=1,~\forall i \in \mathcal{G}_k,~\forall k \label{eqn:gencoh}\\
&x_{i,k}-x_{j,k}\leq y_{i,j}, \forall (i,j) \in \mathcal{E},~\forall k \label{eqn:dc_swchg1}\\\
&x_{j,k}-x_{i,k}\leq y_{i,j}, \forall (i,j) \in \mathcal{E},~\forall k \label{eqn:dc_swchg2}\\
&x_{i,k}+x_{j,k}+y_{i,j} \leq 2, \forall (i,j) \in \mathcal{E},~\forall k \label{eqn:dc_swchg3}\\
&\sum_{(i,j)\in\mathcal{E}} p_{i,j} - \sum_{(j,i)\in\mathcal{E}} p_{j,i} = P_{G,i}^s - P_{GS,i} - \nonumber \\
& P_{L,i}^s + P_{LS,i},~\forall i \in \mathcal{V} \label{eqn:dc_pwrbal}\\
&P_{GS,i}^{min} \leq P_{GS,i} \leq P_{GS,i}^{max},~\forall i \in \mathcal{V} \label{eqn:dc_Pgmax}\\
&P_{LS,i}^{min} \leq P_{LS,i} \leq P_{L,i}^{s},~\forall i \in \mathcal{V} \label{eqn:dc_Plmax}\\
&p_{i,j}\leq p_{i,j}^{max}(1-y_{i,j}),~\forall (i,j) \in \mathcal{E} \label{eqn:dc_Pij_Hi}\\
&p_{i,j}\geq -p_{i,j}^{max}(1-y_{i,j}),~\forall (i,j) \in \mathcal{E} \label{eqn:dc_Pij_Lo}\\
&p_{i,j}-b_{i,j}(\varphi_i-\varphi_j)\leq M_{i,j}^{\varphi}y_{i,j},~\forall (i,j) \in \mathcal{E} \label{eqn:dc_KVL_Hi}\\
&p_{i,j}-b_{i,j}(\varphi_i-\varphi_j)\geq - M_{i,j}^{\varphi}y_{i,j},~\forall (i,j) \in \mathcal{E} \label{eqn:dc_KVL_Lo}\\
&\varphi^{min} \leq \varphi_i \leq \varphi^{max},~\forall i\in\mathcal{V} \label{eqn:dc_phiMax}\\
&\varphi_r = 0,~r \in \mathcal{R} \label{eqn:dc_phiRef}\\
&f_{i,j}\leq (n-1)(1-y_{i,j}),~\forall (i,j) \in \mathcal{E} \label{eqn:fij_Hi}\\
&f_{i,j}\geq -(n-1)(1-y_{i,j}),~\forall (i,j) \in \mathcal{E} \label{eqn:fij_Lo}\\
&\sum_{(r_k,j)\in\mathcal{E}}f_{r_k,j}-\sum_{(j,r_k)\in\mathcal{E}}f_{j,r_k}=\sum_{i \in \mathcal{V}} x_{i,k},~ \nonumber \\ &r_k \in \mathcal{R},~k=1,\ldots,K,~x_{r_k,k}=1 \label{eqn:fij_srce}\\
&\sum_{(i,j)\in\mathcal{E}}f_{i,j}-\sum_{(j,i)\in\mathcal{E}}f_{j,i}=-1,~\forall i\in\mathcal{V}\setminus\mathcal{R} \label{eqn:fij_sink}\\
&x_{i,k}\in \{0,1\},~\forall i\in \mathcal{V},~k=1,\ldots, K\\
&y_{i,j}\in \{0,1\},~\forall (i,j)\in\mathcal{E}  \label{eqn:XYbin}
\end{align}\label{eqn:dc_opf}
\end{subequations}  

The objective of \eqref{eqn:dc_opf} combines the relevant objectives from \cite{Fan.2012,Kyriacou.2017,Patsakis.2019} with different weight factors. The factor $\beta$ is associated with minimal load shedding, which is of great importance for ICI. While minimizing load shedding, it is important to avoid arbitrary large values of shed or disconnected generation, which is achieved by introducing the minimization of generation shedding into \eqref{eqn:objective1} with the weight $\gamma$. Additionally, \cite{Kyriacou.2017,Patsakis.2019} use the minimization of total power flow disruption as their cost function. In our experience, this choice of cost function allows for an easier convergence to the optimum, but it cannot  well enough represent the power balance within islands, which is the main challenge of ICI. However, a large power flow disruption is undesirable \cite{Ding.2013}, which explains its  presence in \eqref{eqn:objective1} with the weight $\mu$. 

Switching constraints \eqref{eqn:dc_xik}--\eqref{eqn:dc_swchg3} ensure the separation of buses belonging to different islands from each other through open branches ($y_{i,j}=1$). Here \eqref{eqn:dc_swchg3} ensures that lines connecting buses in the same island are closed, thus precluding line switching inside islands. Because of \eqref{eqn:dc_swchg3}, the formulation in \eqref{eqn:dc_opf} becomes easier to solve, as its feasible region becomes more constrained. From an operational perspective, requiring both the optimal splitting cutset and optimal line switching inside of each island may be too complex and impractical during an emergency condition. 

DC OPF constraints \eqref{eqn:dc_pwrbal}--\eqref{eqn:dc_KVL_Lo} describe the physics of DC power flow. Kirchhoff current law (KCL) is modeled by \eqref{eqn:dc_pwrbal}, which represents the power balance at each node. Constraints \eqref{eqn:dc_Pij_Hi}--\eqref{eqn:dc_KVL_Lo} model Ohm's law under the presence of switching decisions $y_{i,j}$, which requires the introduction of the node potential variables $\varphi_i,~\forall i\in\mathcal{V}$. When branch $(i,j)$ is closed, $p_{i,j}$ is governed by Ohm's law. If branch $(i,j)$ is open, $p_{i,j}=0$ and the angle difference $(\varphi_i-\varphi_j)$ can be arbitrary. Therefore, the big-M constants $M_{i,j}^{\varphi}$ are needed to allow $(\varphi_i-\varphi_j)$ to take some rather large values. Additionally, the lower and upper bounds $\varphi^{min}$ and $\varphi^{max}$ in \eqref{eqn:dc_phiMax} can also be seen as big-M constants, as their values often cannot be estimated exactly. 

Single commodity flow constraints are used to ensure the connectedness of each island in the same way as the switching constraints \eqref{eqn:dc_xik}-\eqref{eqn:dc_swchg2} ensure the separation of islands from one another. For each coherent group $\mathcal{G}_k$, a root node $r_k\in\mathcal{G}_k,~r_k\in\mathcal{R}$ is selected, whereby $\mathcal{R}$ is the union of root nodes of all groups. Constraint \eqref{eqn:fij_srce} requires that the number of units of an artificial commodity produced at $r_k$ equals the total number of nodes in island $k$ corresponding to $\mathcal{G}_k$, while constraints \eqref{eqn:fij_sink} demand one unit of commodity to be consumed at each non-root node. The satisfaction of flow balance  \eqref{eqn:fij_srce}--\eqref{eqn:fij_sink} requires each island to be connected. However, the upper bounds of commodity flows for each branch are not known, which prompts the usage of another big-M constant $n-1$ in \eqref{eqn:fij_Hi}--\eqref{eqn:fij_Lo}. This big-M constant could be somewhat lowered through exact calculations, but $n-1$ is shown here to simplify the representation. Unlike \cite{Fan.2012,Kyriacou.2017}, \eqref{eqn:dc_opf} does not require a separate set of commodity flow variables for each island, as it appears that a single set of $m$ commodity flows $f_{ij}$ can be used for all islands.

\section{Proposed MILP Model}
\label{sec:milp_our}
\subsection{DC OPF ICI using Cycle-based KVL Constraints}
\label{sec:DCICIKVL}
A major issue with \eqref{eqn:dc_opf} is the presence of big-M constants $M_{i,j}^{\varphi}$ in \eqref{eqn:dc_KVL_Hi}--\eqref{eqn:dc_KVL_Lo}. While there are methods to bound these coefficients when the solution is required to contain a single connected component (e.g., \cite{Fattahi.2019}), no bound strengthening method is known for the disconnected networks resulting from ICI. Thus, a possibly unrealistically large value needs to be assumed for $M_{i,j}^{\varphi}$ in order to preserve all feasible solutions. 

An alternative to Ohm's law form of power flow equations in \eqref{eqn:dc_KVL_Hi}--\eqref{eqn:dc_KVL_Lo} is to apply KVL to the entire network, while considering the switching decisions. The KVL equations are written with respect to a set of linearly independent cycles in the network, which is known as \emph{cycle basis}. For the reasons that become clear below, it is desirable to compute the cycle basis as quickly as possible, which is achieved by using as cycle basis \emph{fundamental cycles} induced by the network's minimal spanning tree (MST) \cite{Tyuryukanov.2018}. Given cycle basis $\mathcal{CB}$, the KVL alternatives of \eqref{eqn:dc_KVL_Hi}--\eqref{eqn:dc_KVL_Lo} can be formulated as follows:
\begin{subequations}
\begin{align}
&\sum_{(i,j)\in C} \frac{p_{i,j}}{b_{i,j}}\text{sgn}(i,j)\leq\frac{1}{2}M_C\sum_{(i,j)\in C} y_{i,j},~\forall C\in\mathcal{CB} \label{eqn:KVL_Y_Hi} \\
&\sum_{(i,j)\in C} \frac{p_{i,j}}{b_{i,j}}\text{sgn}(i,j)\geq-\frac{1}{2}M_C\sum_{(i,j)\in C} y_{i,j},~\forall C\in\mathcal{CB} \label{eqn:KVL_Y_Lo}
\end{align}\label{eqn:KVL_Y}
\end{subequations}

where $C\in\mathcal{CB}$ is a cycle formed by undirected edges $(i,j)\in\mathcal{E}$, and function $\text{sgn}(i,j)$ takes the plus sign if edge $(i,j)$ is aligned with the clockwise direction of $C$ and the minus sign otherwise. 

As \eqref{eqn:dc_swchg3} precludes line switching inside islands, each cycle is either connected or has at least two open lines. Therefore, $M_C$ is the sum of $|\frac{p_{i,j}^{max}}{b_{i,j}}|, (i,j)\in C$ minus the two smallest values of $|\frac{p_{i,j}^{max}}{b_{i,j}}|$ along $C$. Unlike $M_{i,j}^{\varphi}$, $M_C$ can be computed exactly, but it may still be quite large for long cycles. Thus, to additionally strengthen the model, we are adding some additional KVL constraints \eqref{eqn:KVL_Y} associated with cycles shorter than a certain length. All such cycles can be found in polynomial time by applying to each node techniques based on recursive node neighborhood traversal. Short cycles (e.g., up to length 7) can be found quickly before the start of the MIP solver, but enumerating longer cycles becomes increasingly computationally inefficient. 

Obviously, any cycle basis of the disconnected network contains less cycles than the cycle basis of the original network. As only a small fraction of edges is opened for ICI, the majority of fundamental cycles of the disconnected network coincide with the initial fundamental cycles. However, in some cases the original fundamental cycles do not fully describe the cycle basis of the disconnected network. Thus, for each integer solution obtained during the solution process, cycles violating \eqref{eqn:KVL_Y} need to be identified, and the corresponding inequalities \eqref{eqn:KVL_Y} added to the model by using MIP solver callbacks. A possible computationally efficient way to achieve this is as follows:

\begin{enumerate}
\item Compute the MST of the original network and the associated fundamental cycle basis $\mathcal{CB}_0$.
\item Assign close-to-zero weights to the network edges belonging to the MST.
\item Once an integer solution is found, compute its minimal spanning forest and the associated fundamental cycle basis $\mathcal{CB}_i$ ($i$ is the solution number). The previously assigned small edge weights should promote the alignment of the newly computed cycle basis with the original one.
\item Check the solution for conformity with \eqref{eqn:KVL_Y} (i.e., $\sum_{(i,j)\in C} \frac{p^{*}_{i,j}}{b_{i,j}}\text{sgn}(i,j) = 0,~\forall C\in\mathcal{CB}_i$).
\item If any cycles in $\mathcal{CB}_i$ violate \eqref{eqn:KVL_Y}, use them to add new KVL constraints \eqref{eqn:KVL_Y} to the model.
\end{enumerate}

Besides strengthening the model, the abovementioned KVL constraints based on extra short cycles also improve the satisfaction of \eqref{eqn:KVL_Y} for the intermediate integer feasible solutions.

\subsection{Island Connectivity Constraints}
\label{sec:CONNECT}
The single commodity flow constraints that are most commonly used to enforce islands' contiguity \cite{Fan.2012,Kyriacou.2017,Teymouri.2017} have the drawback of introducing at least $m$ auxiliary variables and large big-M constants that may lead to loose LP relaxations. In \cite{Patsakis.2019}, two alternative methods based on multicommodity flows and network cutsets are proposed. However, these alternatives may often underperform the simple single commodity flow approach. At least the multicommodity flow method has largely theoretical importance due to the very large number of auxiliary variables that it requires \cite{Chopra.2001}. The cutset-based method requires solving many max-flow min-cut problems to identify cutsets that violate connectivity, which has the time complexity of $O(mn^2)$ for a single max-flow min-cut run. 

In this section, a different approach to islands' connectivity is proposed that does not involve big-M coefficients and allows to quickly find strong inequalities that ensure network connectivity. This approach is related to the studies on radiality constraints in distribution networks (e.g., \cite{Borghetti.2012,Jabr.2013}), but has some application-specific features and computational improvements. One of conditions for network connectedness is that there exists a spanning tree with $n-1$ edges that spans all of its nodes. In general, for a network with $k$ connected components there exists $k$ spanning trees (one for each component) forming a \emph{spanning forest} that includes $n-k$ edges. This observation motivates the following connectivity formulation based on rooted directed spanning trees:

\begin{subequations}
\begin{align}
& \sum_{(i,j)\in\mathcal{A}} z_{i,j} = n-k \label{eqn:sptree} \\
& z_{j,r} = 0,~\forall (j,r)\in\mathcal{A},~\forall r \in \mathcal{R} \label{eqn:rootIN} \\
& \sum_{(r,j)\in\mathcal{A}} z_{r,j} \geq 1,~\forall r \in \mathcal{R} \label{eqn:rootOUT} \\
& \sum_{(j,i)\in\mathcal{A}} z_{j,i} = 1,~\forall i \in \mathcal{V}\setminus\mathcal{R} \label{eqn:nodeIN} \\
& z_{i,j}+z_{j,i} \leq 1-y_{i,j},~\forall (i,j)\in \mathcal{E}_C \label{eqn:cyclZ2Y}\\
& z_{i,j}+z_{j,i} = 1-y_{i,j},~\forall (i,j)\in \mathcal{E}\setminus\mathcal{E}_C \label{eqn:treeZ2Y} \\
& \sum_{(i,j)\in C} z_{i,j} \leq |C|-1,~\forall C\in \mathcal{D} \label{eqn:setpack} \\
& z_{i,j}\in \{0,1\},~\forall (i,j)\in\mathcal{A} \label{eqn:z01}
\end{align}\label{eqn:netconn}
\end{subequations} 

where $z_{i,j}$ and $z_{j,i}$ represent the status of arcs $(i,j)$ and $(j,i)\in\mathcal{A}$ ($z_{i,j}=1$ iff $(i,j)$ is enabled). 

The equalities in \eqref{eqn:rootIN}--\eqref{eqn:rootOUT} define the root nodes of $K$ coherent generator groups as root nodes of $K$ trees each spanning one island. According to \eqref{eqn:rootIN}--\eqref{eqn:rootOUT}, each root node should have no incoming arcs and at least one outgoing arc. Conversely, each non-root node should have one incoming arc according to \eqref{eqn:nodeIN}. Constraints \eqref{eqn:cyclZ2Y}--\eqref{eqn:treeZ2Y} link $z$ and $y$ variables. For the edges that participate in network cycles (i.e., in the cycle basis described in Section \ref{sec:DCICIKVL}), \eqref{eqn:cyclZ2Y} is valid. For the rest of the edges, the stronger version \eqref{eqn:treeZ2Y} can be used instead of  \eqref{eqn:cyclZ2Y}. The number of  enabled arcs is constrained by \eqref{eqn:sptree}. 


Combined constraints \eqref{eqn:sptree}--\eqref{eqn:treeZ2Y} can define $k$ connected components spanned by $k$ directed trees, or more than $k$ connected components. In the latter case, the additional components will form directed cycles to satisfy \eqref{eqn:sptree}--\eqref{eqn:treeZ2Y}. Therefore, cycle breaking constraints \eqref{eqn:setpack} are needed to only allow spanning forests as solutions, where $\mathcal{D}$ is the set of all directed cycles in the network and $C$ is a sequence of arcs in $\mathcal{A}$ forming a directed cycle. As there are exponentially many cycles in the network, all constraints in \eqref{eqn:setpack} cannot be simultaneously added to the model. Instead, only the violated inequalities \eqref{eqn:setpack} are added in MIP solver callbacks that are available in most of modern MIP solvers  (see Section \ref{sec:mipsolve}).

\subsection{Extra Strengthening Cycle Inequalities}
\label{sec:cycle}
To further strengthen the cycle-based ICI formulation, the following \emph{triangle inequalities} (e.g., see \cite{Hojny.2021}) are added for all known cycles of the minimal length:
\begin{equation}
y_{i,j} - \sum_{\substack{(p,q)\in C\\ (p,q)\neq(i,j)}} y_{p,q} \leq 0,~\forall C\in \mathcal{CB}, |C|=3, \forall (i,j)\in C 
\label{eqn:triang}
\end{equation}

\subsection{Minimization of Load-Generation Imbalance}
\label{sec:IMBLNCE}
Minimal load shedding is often chosen as the cost function of ICI, as it is highly important for utilities to minimize the loss of load. However, shedding an exact amount of load at exact bus at exact moment in time to realize the outcome of \eqref{eqn:dc_opf} may be sometimes quite challenging with the available equipment. Moreover, the optimal load shedding can be zero, in which case it will be impossible to provide a non-zero lower bound, and the integrality gap of the MIP solver will be close to 100~\% during the whole optimization. 

Among practitioners, the notion of \emph{balanced islands} with good load-generation balance is no less popular than the idea of minimal load shedding. Islands with good power balance would naturally limit load shedding without the need of picking specific buses, while also limiting the ROCOF value after splitting. At the same time, the total power imbalance of all islands is always non-zero and can be much easier bounded from below. Based on \eqref{eqn:dc_opf}, the minimization of load generation imbalance can be formulated as follows:

\begin{subequations}
\begin{align}
&\text{min} \quad \sum_{k=1}^K P_{\Delta,k} \label{eqn:objIMBLOD} \\
&\quad P_{\Delta,k} \geq \sum_{i\in\mathcal{V}}(P_{G,i}^s-P_{L,i}^s)x_{i,k},~ \forall k \label{eqn:genimb}\\
&P_{\Delta,k} \geq \sum_{i\in\mathcal{V}}(P_{L,i}^s-P_{G,i}^s)x_{i,k},~\forall k \label{eqn:lodimb}\\
&\eqref{eqn:dc_xik}-\eqref{eqn:XYbin} \nonumber
\end{align}\label{eqn:IMBLOD}
\end{subequations}  

where $P_{\Delta,k}$ represents the power imbalance of island $k$. 

The inequalities in \eqref{eqn:genimb}--\eqref{eqn:lodimb} are constraining the absolute value $|\sum_{i\in\mathcal{V}}(P_{G,i}^s-P_{L,i}^s)x_{i,k}|$ from below for each $k$, while the minimization requirement \eqref{eqn:objIMBLOD} constrains it from above. 

\subsection{Overall Solution Process}
\label{sec:mipsolve}
The ideas from Sections \ref{sec:CONNECT}--\ref{sec:DCICIKVL} can be combined together into the following novel DC OPF ICI formulation: 
\begin{align}
\nonumber
&\text{min}~\alpha\sum_{k=1}^K P_{\Delta,k} + \beta\sum_{i\in\mathcal{V}}P_{LS,i} + \gamma\sum_{i\in\mathcal{V}}P_{GS,i} + \mu\sum_{(i,j)\in\mathcal{E}}p_{i,j}^s y_{i,j} \\\nonumber
&\eqref{eqn:dc_xik}-\eqref{eqn:dc_Pij_Lo} \\\label{eqn:dc_our}
&\eqref{eqn:KVL_Y_Hi}-\eqref{eqn:KVL_Y_Lo} \\\nonumber
&\eqref{eqn:sptree}-\eqref{eqn:z01},~\eqref{eqn:triang},~\eqref{eqn:genimb}-\eqref{eqn:lodimb} \nonumber
\end{align} 

The above formulation requires MILP callbacks for cycle breaking constraints \eqref{eqn:setpack} to be solved efficiently. In our implementation, the violated cycle breaking constraints are easily separated using Algorithm~\ref{alg:cycbreak} for each new integer solution.
\begin{algorithm}[t]
\begin{algorithmic}[1]
\renewcommand{\algorithmiccomment}[1]{\small//~#1\normalsize} 
\Statex \textbf{Input:} $z_{i,j}^*$, $\mathcal{V}$, $K$
\Statex \textbf{Output:} Set of violating directed cycles $\mathcal{C}$
\State $G(\mathcal{V},\mathcal{A}_G)\gets \text{directed graph induced by~}z_{i,j}^*$
\State $F(\mathcal{V},\mathcal{A}_F)\gets\mathtt{spanning\_forest}(G)$
\State $\mathcal{CC}\gets \mathtt{connected\_components}(F)$
\If {$|\mathcal{CC}|=K$}
\Return $\emptyset$
\EndIf
\State $\mathcal{A}_C\gets \mathcal{A}_G\setminus\mathcal{A}_F$
\For {$a_{i,j} \in \mathcal{A}_C$}
\State $P\gets \mathtt{shortest\_path}(\mathcal{F},\text{source=}j,\text{target=}i)$
\State $C\gets P\cup a_{i,j}$
\State $\mathcal{C}\gets\mathcal{C}\cup C$
\EndFor\\
\Return $\mathcal{C}$
\end{algorithmic}
\caption{Generation of cycle breaking constraints}
\label{alg:cycbreak}
\end{algorithm}

Given an integer solution $z_{i,j}^*$, it can be considered as directed graph $G$. For this graph, the spanning forest $F$ can be computed by using the standard spanning tree algorithms with loglinear time complexity. If $F$ has $k$ connected components, the solution is accepted, and the best integer solution gets updated. Otherwise, Algorithm \ref{alg:cycbreak} finds arcs that are in $G$, but not in $F$. Each such arc corresponds to a directed cycle, which allows to retrieve the full set of cycles $\mathcal{C}$ that violate \eqref{eqn:setpack} in loglinear time.

In addition to cycle breaking constraints \eqref{eqn:setpack}, we are adding directed cutset constraints  between the root node of each group and the remaining generators of the same group using the same $z_{i,j}$ variables. These constraints are convenient for enforcing connectivity of a defined set of \emph{terminal nodes} to the root node. A detailed description of these constraints including improvements like \emph{back cuts} and \emph{creep flow}  can be found in \cite{Chopra.2001} and several other references. 

The cycle breaking constraints in \eqref{eqn:setpack} also combine well with the cycle-based DC OPF ICI formulation in \eqref{eqn:KVL_Y}. To apply \eqref{eqn:KVL_Y}, a cycle basis and optionally some additional network cycles need to be found. From these precomputed cycles, the shorter ones (e.g., with the length less than 6--8 edges) can be used to strengthen the initial model formulation with some constraints of type \eqref{eqn:setpack}. 

\section{MILP Heuristics}
\label{sec:milp_heur}
When solving \eqref{eqn:dc_our}, no feasible solution could be found within the optimization time limit for several larger instances. In practice, it is highly beneficial to obtain an initial feasible solution as early as possible because this would increase the available time to improve the upper bound of \eqref{eqn:dc_our} using various efficient techniques available in the modern MIP solvers. As the practice has shown that multiple MILP heuristics available in Gurobi \cite{Gurobi.2020} often cannot retrieve a feasible solution, a new problem-specific heuristic had to be proposed to improve the solvability of \eqref{eqn:dc_opf} and \eqref{eqn:dc_our}. This heuristic is outlined in Algorithm~\ref{alg:mipheur}.
\begin{algorithm}[t]
\begin{algorithmic}[1]
\renewcommand{\algorithmiccomment}[1]{\small//~#1\normalsize} 
\Statex \textbf{Input:} $x_{i,k}^{LP}$, $\mathcal{V}$, $\mathcal{E}$, $\mathcal{R}$, K
\Statex \textbf{Output:} $x_{i,k}^{*}$, $y_{i,j}^{*}$, $z_{i,j}^{*}$
\State $\mathcal{E}^{on}\gets\emptyset$
\For{$k=1,\ldots,K$}  \label{lin:EONstrt}
\For{$(i,j)\in\mathcal{E}$}
\If{$x_{i,k}^{LP}>0.9\wedge x_{j,k}^{LP}>0.9$} $\mathcal{E}^{on}\gets\mathcal{E}^{on}\cup (i,j)$
\EndIf
\EndFor
\EndFor  \label{lin:EONend}
\State $\mathcal{CC} \gets \mathtt{connected\_components}(G(\mathcal{V},\mathcal{E}^{on}))$ \label{lin:GCC}
\State $\mathcal{P} \gets \emptyset$ \Comment{Set of partial islands}
\For{$CC\in\mathcal{CC}$}
\State $r \gets\mathcal{R}\cap CC$
\If{$|r|=0$} \textbf{continue}
\ElsIf{$|r|\geq 2$}
\Return $\emptyset$
\Else $~\mathcal{P}\gets \mathcal{P}\cup CC$
\EndIf
\EndFor
\If{$\frac{\sum_{P\in\mathcal{P}}|P|}{n}\geq0.8$}
\State \parbox[t]{\dimexpr\linewidth-\algorithmicindent}{Fix $x_{i,k}$, $y_{i,j}$, $z_{i,j}$ according to $\mathcal{P}$ and solve the reduced MILP (e.g., \eqref{eqn:dc_opf}) with more than 80\% fixed buses.\strut} 
\EndIf\\
\Return The solution of reduced MILP $x_{i,k}^{*}$, $y_{i,j}^{*}$, $z_{i,j}^{*}$
\end{algorithmic}
\caption{MILP heuristic for initialization of DC OPF ICI}
\label{alg:mipheur}
\end{algorithm}

The LR of $x_{i,k}$ is taken as input and analyzed in lines \ref{lin:EONstrt}--\ref{lin:EONend} to identify the edges that are considered to be closed by the current LR. Then these edges are used to build an undirected graph $G$ in line \ref{lin:GCC}, and the connected components of that graph are identified. Next, each connected component is analyzed for the presence of a root node $r\in\mathcal{R}$ in it. If a root node is present is the component, it is saved to the set of \emph{partial islands} $\mathcal{P}$. Finally, the $x_{i,k}$ variables of the nodes belonging to contiguous partial islands are fixed, and the values of the remaining node variables are obtained by solving the initial problem as MILP with the majority of discrete variables being fixed. Because of a high degree of reduction of discrete variables, the reduced MILP is usually solved in a fraction of a second. However, it is important to find connected partial islands, as just rounding the LR will very often result in infeasibility.

The heuristic in Algorithm~\ref{alg:mipheur} is only run until a feasible solution becomes available, as modern MILP solvers are able to apply more efficient  heuristics to consistently improve the existing solution.

\section{Computational Results}
\label{sec:results}
\subsection{Test Setup}
\label{sec:setup}
This section\footnote{The code implementation can be found at \url{https://github.com/ityuryukanov/power_network_clustering} under \texttt{.papers/PSCC\_2022}.} presents a comparison of the proposed formulation in \eqref{eqn:dc_our} to the known benchmark model in \eqref{eqn:dc_opf}. The results are produced for a number of test networks from the MATPOWER toolbox \cite{Fliscounakis.2013}. As publicly available power system test cases do not include coherent generator groupings that are required in \eqref{eqn:gencoh}, those were obtained with the generator coherency algorithm from \cite{Tyuryukanov.2020} for every test case. For the MATPOWER test cases lacking the dynamic generator data it was fitted from \cite{Anderson.2003}.

Three test cases from MATPOWER were selected to evaluate the proposed formulation: \emph{case89pegase}, \emph{case1354pegase}, and \emph{case1888rte}. The 89 bus test case was selected because it posed some difficulties for the MILP solver despite its moderate size, while the 1354 and 1888 bus test cases are examples of large-scale networks.

All computations were performed using Gurobi v.~9.1. \cite{Gurobi.2020} on a PC with an Intel\textsuperscript{\textregistered} Core\textsuperscript{\texttrademark} i7 2.20 GHz CPU with 6 cores and 16~Gb of RAM running on Windows 10. The reported solution times include the presolve and MILP solver time, but exclude the model preparation time, which is polynomial. The time limit for the test cases with less than 500 buses has been set to 480 seconds and to 720 seconds for the test cases with more than 500 buses. In all the experiments, the default Gurobi solver settings were used. The maximal time to be spent in the MILP heuristic from Section \ref{sec:milp_heur} is set to 3 \% of the time limit. For simplicity, the auxiliary MILP model in Algorithm \ref{alg:mipheur} is based on \eqref{eqn:dc_opf} for all test cases.  The solution was considered optimal if its MIP optimality gap was less than 1\%.
\begin{table}[t]
\centering
\caption{Load Shedding Minimization using \eqref{eqn:dc_opf}}
\begin{tabular}{|c|c|c|c|c|c|c|c|c|}\hline
 $n$ & $K$   & $UB$,         & $g,~\%$         & $P_{LS}$,    & $P_{\Delta}$, & $P_{GS}$,    & $p_{ij}^{\Sigma}$,\\
     & ~     & p.u.          & $T$,~s          & p.u.         & p.u.          & p.u.         & p.u. \\\hline
\small{89} & \small{2} &\small{6.4089} &\small{47~\%} &\small{5.02} &\small{11.4} &\small{6.4} &\small{13.3}\\
\small{89} & \small{3} &\small{8.3011} &\small{46~\%} &\small{5.02} &\small{11.4} &\small{6.4} &\small{32.2}\\
\small{89} & \small{4} &\small{13.844} &\small{64~\%} &\small{10.2} &\small{21.9} &\small{11.6} &\small{34.9}\\
\small{89} & \small{5} &\small{13.876} &\small{70~\%} &\small{10.2} &\small{21.9} &\small{11.6} &\small{35.2}\\\hline
\small{1354} & \small{2} &\small{2.3878} &\small{0.5~s} &\small{-0.00} &\small{16.6} &\small{16.6} &\small{22.2}\\
\small{1354} & \small{3} &\small{4.6645} &\small{100~s} &\small{-0.00} &\small{16.6} &\small{16.6} &\small{45}\\
\small{1354} & \small{4} &\small{3.3921} &\small{1~s} &\small{-0.00} &\small{16.6} &\small{16.6} &\small{32.3}\\
\small{1354} & \small{5} &\small{8.3761} &\small{20~\%} &\small{-0.00} &\small{16.6} &\small{16.6} &\small{82.1}\\
\small{1354} & \small{6} &\small{12} &\small{12~\%} &\small{0.03} &\small{16.7} &\small{16.7} &\small{118}\\
\small{1354} & \small{7} &\small{13.631} &\small{14~\%} &\small{0.01} &\small{16.7} &\small{16.6} &\small{135}\\
\small{1354} & \small{8} &\small{21.148} &\small{33~\%} &\small{1.58} &\small{19.8} &\small{18.2} &\small{194}\\\hline
\small{1888} & \small{2} &\small{0.4949} &\small{0.8~s} &\small{-0.00} &\small{9.81} &\small{9.81} &\small{3.97}\\
\small{1888} & \small{3} &\small{11.38} &\small{51~\%} &\small{0.08} &\small{9.81} &\small{9.88} &\small{112}\\
\small{1888} & \small{4} &\small{4.5363} &\small{11~\%} &\small{-0.00} &\small{9.81} &\small{9.81} &\small{44.4}\\
\small{1888} & \small{5} &\small{7.0738} &\small{14~\%} &\small{-0.00} &\small{9.81} &\small{9.81} &\small{69.8}\\
\small{1888} & \small{6} &\small{9.2755} &\small{10~\%} &\small{0.02} &\small{9.85} &\small{9.83} &\small{91.5}\\
\small{1888} & \small{7} &\small{16.976} &\small{40~\%} &\small{0.17} &\small{9.94} &\small{9.97} &\small{167}\\
\small{1888} & \small{8} &\small{12.457} &\small{17~\%} &\small{-0.00} &\small{9.81} &\small{9.81} &\small{124}\\\hline
\end{tabular}     
\label{tab:part_time_LS_flow}
\end{table}

\subsection{Minimization of Load Shedding}
\label{sec:totalLS}
The first set of test results is related to minimizing the known objective in \eqref{eqn:objective1}. The objective weights are set as follows: $\alpha=0$, $\beta=1$, $\gamma=0.01$, $\mu=0.1$. The relatively high value of $\mu$ is selected mainly to ensure non-zero lower bounds in all cases and thus a more informative progression of the MIP optimality gap during the optimization. The big-M coefficients $M_{i,j}^{\varphi}$ in \eqref{eqn:dc_our} are set to $2\pi$, $\varphi^{min}$ is set to $-\pi$, $\varphi^{max}$ is set to $\pi$ , and the power flow limitation $p_{ij}^{max}$ is assumed to be equal to $b_{i,j}\pi/4,~\forall (i.j)\in\mathcal{E}$.

First, the results for the baseline formulation in \eqref{eqn:dc_opf} are presented in Table \ref{tab:part_time_LS_flow}. The table header can be read as follows: $UB$ is the best MIP objective value (i.e., the upper bound), $g$ is the final optimality gap, $T$ is the solution time, $P_{LS}$, $P_{\Delta}$, $P_{GS}$, $p_{ij}^{\Sigma}$ are the final values of load shedding, total power imbalance, generator shedding, and power flow cut respectively. In Table \ref{tab:part_time_LS_flow}, $g$ and $T$ are included interchangeably: if $g$ is less than the optimality tolerance then $T$ is included, otherwise $g$ is shown, as $T$ equals to the time limit.  

Next, in the formulation in \eqref{eqn:dc_opf} the flow-based connectivity constraints  \eqref{eqn:fij_Hi}--\eqref{eqn:fij_sink} are replaced with \eqref{eqn:netconn} and the MILP heuristic is enabled. The results obtained after this modification are summarized in Table \ref{tab:part_time_LS_zij}. Finally, the results of our complete formulation in \eqref{eqn:dc_our} with the MILP heuristic enabled for the given cost function are given in Table \ref{tab:part_time_LS_our}. 

As it can be seen, the results in Tables \ref{tab:part_time_LS_zij}--\ref{tab:part_time_LS_our} in terms of the MIP optimality gap and solution time usually exceed those in Table \ref{tab:part_time_LS_flow}, which demonstrates the efficacy of the proposed improvements. 

\begin{table}[t]
\centering
\caption{Load Shedding Minimization using \eqref{eqn:dc_opf} and \eqref{eqn:netconn}}
\begin{tabular}{|c|c|c|c|c|c|c|c|c|}\hline
 $n$ & $K$   & $UB$,   	  & $g,~\%$  	  & $P_{LS}$,    & $P_{\Delta}$, & $P_{GS}$,    & $p_{ij}^{\Sigma}$,\\
	 & ~     & p.u.    	  & $T$,~s   	  & p.u.         & p.u.          & p.u.         & p.u.\\\hline
\small{89} & \small{2} &\small{6.4089} &\small{49~\%} &\small{5.02} &\small{11.4} &\small{6.4} &\small{13.3}\\
\small{89} & \small{3} &\small{8.3011} &\small{46~\%} &\small{5.02} &\small{11.4} &\small{6.4} &\small{32.2}\\
\small{89} & \small{4} &\small{13.844} &\small{65~\%} &\small{10.2} &\small{21.9} &\small{11.6} &\small{34.9}\\
\small{89} & \small{5} &\small{13.876} &\small{68~\%} &\small{10.2} &\small{21.9} &\small{11.6} &\small{35.2}\\\hline
\small{1354} & \small{2} &\small{2.3878} &\small{0.5~s} &\small{-0.00} &\small{16.6} &\small{16.6} &\small{22.2}\\
\small{1354} & \small{3} &\small{4.6645} &\small{80~s} &\small{-0.00} &\small{16.6} &\small{16.6} &\small{45}\\
\small{1354} & \small{4} &\small{3.3921} &\small{1.1~s} &\small{-0.00} &\small{16.6} &\small{16.6} &\small{32.3}\\
\small{1354} & \small{5} &\small{8.1682} &\small{16~\%} &\small{-0.00} &\small{16.6} &\small{16.6} &\small{80}\\
\small{1354} & \small{6} &\small{12} &\small{12~\%} &\small{0.03} &\small{16.7} &\small{16.7} &\small{118}\\
\small{1354} & \small{7} &\small{13.453} &\small{13~\%} &\small{0.03} &\small{16.7} &\small{16.7} &\small{133}\\
\small{1354} & \small{8} &\small{20.278} &\small{30~\%} &\small{0.29} &\small{17.2} &\small{16.9} &\small{198}\\\hline
\small{1888} & \small{2} &\small{0.4949} &\small{1~s} &\small{-0.00} &\small{9.81} &\small{9.81} &\small{3.97}\\
\small{1888} & \small{3} &\small{9.0864} &\small{39~\%} &\small{0.01} &\small{9.82} &\small{9.81} &\small{89.8}\\
\small{1888} & \small{4} &\small{4.5119} &\small{10~\%} &\small{-0.00} &\small{9.81} &\small{9.81} &\small{44.1}\\
\small{1888} & \small{5} &\small{7.0672} &\small{14~\%} &\small{-0.00} &\small{9.81} &\small{9.81} &\small{69.7}\\
\small{1888} & \small{6} &\small{9.3695} &\small{12~\%} &\small{-0.00} &\small{9.81} &\small{9.81} &\small{92.7}\\
\small{1888} & \small{7} &\small{14.329} &\small{28~\%} &\small{-0.00} &\small{9.81} &\small{9.81} &\small{142}\\
\small{1888} & \small{8} &\small{12.158} &\small{15~\%} &\small{0.00} &\small{9.81} &\small{9.81} &\small{121}\\\hline
\end{tabular}
\label{tab:part_time_LS_zij}
\end{table}

\begin{table}[t]
\centering
\caption{Load Shedding Minimization using \eqref{eqn:dc_our}}
\begin{tabular}{|c|c|c|c|c|c|c|c|c|}\hline
 $n$ & $K$ & $UB$,        & $g,~\%$        & $P_{LS}$,    & $P_{\Delta}$, & $P_{GS}$,    & $p_{ij}^{\Sigma}$,\\
	 & ~   & p.u.    	  & $T$,~s   	   & p.u.         & p.u.          & p.u.         & p.u.\\\hline
\small{89} & \small{2} &\small{6.4089} &\small{57~\%} &\small{5.02} &\small{11.4} &\small{6.4} &\small{13.3}\\
\small{89} & \small{3} &\small{8.3011} &\small{57~\%} &\small{5.02} &\small{11.4} &\small{6.4} &\small{32.2}\\
\small{89} & \small{4} &\small{11.254} &\small{64~\%} &\small{5.9} &\small{13.2} &\small{7.28} &\small{52.9}\\
\small{89} & \small{5} &\small{11.096} &\small{62~\%} &\small{5.9} &\small{13.2} &\small{7.28} &\small{51.3}\\\hline
\small{1354} & \small{2} &\small{2.3878} &\small{0.5~s} &\small{-0.00} &\small{16.6} &\small{16.6} &\small{22.2}\\
\small{1354} & \small{3} &\small{4.6645} &\small{55~s} &\small{-0.00} &\small{16.6} &\small{16.6} &\small{45}\\
\small{1354} & \small{4} &\small{3.3921} &\small{1.1~s} &\small{-0.00} &\small{16.6} &\small{16.6} &\small{32.3}\\
\small{1354} & \small{5} &\small{8.1682} &\small{17~\%} &\small{-0.00} &\small{16.6} &\small{16.6} &\small{80}\\
\small{1354} & \small{6} &\small{12} &\small{11~\%} &\small{0.03} &\small{16.7} &\small{16.7} &\small{118}\\
\small{1354} & \small{7} &\small{13.441} &\small{13~\%} &\small{-0.00} &\small{16.6} &\small{16.6} &\small{133}\\
\small{1354} & \small{8} &\small{18.420} &\small{21~\%} &\small{0.24} &\small{17.1} &\small{16.9} &\small{180}\\\hline
\small{1888} & \small{2} &\small{0.4949} &\small{0.6~s} &\small{-0.00} &\small{9.81} &\small{9.81} &\small{3.97}\\
\small{1888} & \small{3} &\small{8.5821} &\small{33~\%} &\small{-0.00} &\small{9.81} &\small{9.81} &\small{84.9}\\
\small{1888} & \small{4} &\small{4.5116} &\small{590~s} &\small{-0.00} &\small{9.81} &\small{9.81} &\small{44.1}\\
\small{1888} & \small{5} &\small{7.0534} &\small{9~\%} &\small{-0.00} &\small{9.81} &\small{9.81} &\small{69.6}\\
\small{1888} & \small{6} &\small{9.275} &\small{11~\%} &\small{0.02} &\small{9.85} &\small{9.83} &\small{91.5}\\
\small{1888} & \small{7} &\small{14.163} &\small{26~\%} &\small{-0.00} &\small{9.81} &\small{9.81} &\small{141}\\
\small{1888} & \small{8} &\small{12.16} &\small{13~\%} &\small{-0.00} &\small{9.81} &\small{9.81} &\small{121}\\\hline
\end{tabular}
\label{tab:part_time_LS_our}
\end{table}

\subsection{Minimization of Total Power Imbalance}
\label{sec:totalIMB}
To complement Section~\ref{sec:totalLS}, the same test cases are recomputed with the main objective of power imbalance minimization. The objective weights are set as follows: $\alpha=1$, $\beta=0.01$, $\gamma=0.01$, $\mu=0.01$. Now the highest priority is given to generation-load imbalance minimization, with load shedding being included with a small weight. As the initial power imbalance is nearly always substantially higher than zero after the network is split, non-zero lower bounds can be produced without assigning an increased value to $\mu$.

The results comparison follows the same route as in Section \ref{sec:totalLS}. The results of minimizing the new objective with the formulation in \ref{eqn:dc_opf} are listed in Table \ref{tab:part_time_IMB_flow}. The outcome of the exchange of \eqref{eqn:fij_Hi}--\eqref{eqn:fij_sink} with \eqref{eqn:netconn} is illustrated in Table \ref{tab:part_time_IMB_zij}. Finally, the results of our formulation in \eqref{eqn:dc_our} are shown in Table \ref{tab:part_time_IMB_our}. 

As it can be seen from Table \ref{tab:part_time_IMB_flow}, with the change of the objective function more cases fail to compute at least a feasible solution with the basic formulation \eqref{eqn:dc_opf}. Thus, the role of feasibility MILP heuristics becomes larger. In fact, the proposed MILP heuristic is enabled for the results in Tables \ref{tab:part_time_IMB_zij}--\ref{tab:part_time_IMB_our}, and feasibility is achieved for all the test cases. The performance is also improved with the proposed enhancements, which can be seen by comparing the 4\textsuperscript{th} columns of Tables \ref{tab:part_time_IMB_flow}--\ref{tab:part_time_IMB_our}. 

An additional important observation can be made by comparing the upper bounds of \emph{case89pegase} across Tables \ref{tab:part_time_IMB_flow}--\ref{tab:part_time_IMB_our}. Although the optimal solution could be reached in all cases, the optimal value is better with our complete formulation \eqref{eqn:dc_our}. This is explained by the absence of big-M constraints in  \eqref{eqn:dc_our}, the values for which must be selected. Thus, \eqref{eqn:dc_our} can be certain to achieve the true optimal solution. In the present case study, it is possible to achieve the same optimum using \eqref{eqn:dc_opf} if the angular big-M constraints are substantially increased (e.g., by a factor of 10).

\begin{table}[t]
\centering
\caption{Islands' Power Imbalance Minimization using \eqref{eqn:dc_opf}}
\begin{tabular}{|c|c|c|c|c|c|c|c|c|}\hline
 $n$ & $K$ & $UB$,        & $g,~\%$       & $P_{LS}$,    & $P_{\Delta}$,   & $P_{GS}$,    & $p_{ij}^{\Sigma}$,\\
	 & ~   & p.u.         & $T$,~s        & p.u.         & p.u.            & p.u.         & p.u.      \\\hline
\small{89} & \small{2} &\small{7.8621} &\small{5~s} &\small{3.09} &\small{6.24} &\small{4.47} &\small{55.5}\\
\small{89} & \small{3} &\small{8.0531} &\small{33~s} &\small{3.09} &\small{6.24} &\small{4.47} &\small{74.6}\\
\small{89} & \small{4} &\small{14.807} &\small{13~s} &\small{8.77} &\small{11.1} &\small{10.1} &\small{66.6}\\
\small{89} & \small{5} &\small{14.809} &\small{16~s} &\small{8.77} &\small{11.1} &\small{10.1} &\small{66.9}\\\hline
\small{1354} & \small{2} &\small{17.023} &\small{0.5~s} &\small{-0.00} &\small{16.6} &\small{16.6} &\small{22.2}\\
\small{1354} & \small{3} &\small{17.272} &\small{11~s} &\small{-0.00} &\small{16.6} &\small{16.6} &\small{47.1}\\
\small{1354} & \small{4} &\small{17.124} &\small{1.7~s} &\small{-0.00} &\small{16.6} &\small{16.6} &\small{32.3}\\
\small{1354} & \small{5} &\small{17.684} &\small{580~s} &\small{-0.00} &\small{16.6} &\small{16.6} &\small{88.3}\\
\small{1354} & \small{6} &\small{--} &\small{--} &\small{--} &\small{--} &\small{--} &\small{--}\\
\small{1354} & \small{7} &\small{--} &\small{--} &\small{--} &\small{--} &\small{--} &\small{--}\\
\small{1354} & \small{8} &\small{--} &\small{--} &\small{--} &\small{--} &\small{--} &\small{--}\\\hline
\small{1888} & \small{2} &\small{9.9447} &\small{1.2~s} &\small{-0.00} &\small{9.81} &\small{9.81} &\small{3.97}\\
\small{1888} & \small{3} &\small{--} &\small{--} &\small{--} &\small{--} &\small{--} &\small{--}\\
\small{1888} & \small{4} &\small{10.427} &\small{23~s} &\small{-0.00} &\small{9.81} &\small{9.81} &\small{52.2}\\
\small{1888} & \small{5} &\small{10.603} &\small{94~s} &\small{-0.00} &\small{9.81} &\small{9.81} &\small{69.8}\\
\small{1888} & \small{6} &\small{10.896} &\small{120~s} &\small{-0.00} &\small{9.81} &\small{9.81} &\small{99.1}\\
\small{1888} & \small{7} &\small{--} &\small{--} &\small{--} &\small{--} &\small{--} &\small{--}\\
\small{1888} & \small{8} &\small{--} &\small{--} &\small{--} &\small{--} &\small{--} &\small{--}\\\hline
\end{tabular}
\label{tab:part_time_IMB_flow}
\end{table}

\footnote{\noindent This work was financially supported by the Dutch Scientific Council NWO in collaboration with TSO TenneT, DSOs Alliander, Stedin, Enduris, VSL and General Electric in the framework of the Energy System Integration \& Big Data program under the project "Resilient Synchromeasurement-based Grid Protection Platform, no. 647.003.004". The authors also greatly thank the anonymous reviewers for their highly useful suggestions.}
\bibliographystyle{IEEEtran}

\section{Conclusions}
\label{sec:conclu}
This paper has proposed several computational enhancements to the DC OPF ICI problem. The main idea was to eliminate the big-M constants that are present in the most existing formulations in order to tighten the linear relaxation of DC OPF ICI. The big-M constants present in island connectivity constraints have been removed by replacing the popular single commodity flow based connectivity model with the new model based on directed spanning forests, which does not require any large coefficients. The big-M constants associated with ICI switching decisions and DC OPF in general have been excluded by replacing the DC OPF switching constraints based on Ohm's law with the new ones based on KVL. 

Besides of handling of all the three types of big-M constants in the standard DC OPF ICI formulation, it has also been observed that the existing ICI objective functions may not fully correspond to the practical requirements associated with system splitting. To ameliorate this, the new objective function minimizing the load-generation imbalance after splitting has been introduced. With this objective, situations when the MIP solver cannot find an initial feasible solution have become more common. To fix this, a new MILP heuristic for DC OPF ICI has been proposed as well. In general, the solution process was noticeably influenced by the type of objective function. Thus, it could be possible to choose "convenient" objective functions to quickly find feasible initial solutions. 

Finally, the proposed enhancements have been tested against the compact and efficient baseline DC ICI OPF model from the literature, and it has been found that the novel ICI model produces better results in most of cases. In the future, it is planned to apply the findings of this paper to more complex and accurate ICI models and possibly to other transmission switching problems.

\begin{table}[t]
\centering
\caption{Power Imbalance Minimization using \eqref{eqn:dc_opf} and \eqref{eqn:netconn}}
\begin{tabular}{|c|c|c|c|c|c|c|c|c|}\hline
 $n$ & $K$ & $UB$,        & $g,~\%$       & $P_{LS}$,   & $P_{\Delta}$,   & $P_{GS}$,    & $p_{ij}^{\Sigma}$,\\
	 &  ~  & p.u.         & $T$,~s        & p.u.        & p.u.            & p.u.         & p.u.\\\hline
\small{89} & \small{2} &\small{7.8621} &\small{13~s} &\small{3.09} &\small{6.24} &\small{4.47} &\small{55.5}\\
\small{89} & \small{3} &\small{8.0531} &\small{85~s} &\small{3.09} &\small{6.24} &\small{4.47} &\small{74.6}\\
\small{89} & \small{4} &\small{14.807} &\small{13~s} &\small{8.77} &\small{11.1} &\small{10.1} &\small{66.6}\\
\small{89} & \small{5} &\small{14.809} &\small{15~s} &\small{8.77} &\small{11.1} &\small{10.1} &\small{66.9}\\\hline
\small{1354} & \small{2} &\small{17.023} &\small{0.6~s} &\small{-0.00} &\small{16.6} &\small{16.6} &\small{22.2}\\
\small{1354} & \small{3} &\small{17.32} &\small{19~s} &\small{-0.00} &\small{16.6} &\small{16.6} &\small{51.9}\\
\small{1354} & \small{4} &\small{17.123} &\small{1.5~s} &\small{-0.00} &\small{16.6} &\small{16.6} &\small{32.3}\\
\small{1354} & \small{5} &\small{17.72} &\small{90~s} &\small{0.07} &\small{16.6} &\small{16.7} &\small{89.6}\\
\small{1354} & \small{6} &\small{18.001} &\small{220~s} &\small{-0.00} &\small{16.6} &\small{16.6} &\small{120}\\
\small{1354} & \small{7} &\small{18.162} &\small{410~s} &\small{-0.00} &\small{16.6} &\small{16.6} &\small{136}\\
\small{1354} & \small{8} &\small{18.628} &\small{1.1~\%} &\small{0.00} &\small{16.6} &\small{16.6} &\small{183}\\\hline
\small{1888} & \small{2} &\small{9.9447} &\small{1~s} &\small{-0.00} &\small{9.81} &\small{9.81} &\small{3.97}\\
\small{1888} & \small{3} &\small{11.135} &\small{3.5~\%} &\small{-0.00} &\small{9.81} &\small{9.81} &\small{123}\\
\small{1888} & \small{4} &\small{10.374} &\small{36~s} &\small{-0.00} &\small{9.81} &\small{9.81} &\small{46.9}\\
\small{1888} & \small{5} &\small{10.688} &\small{280~s} &\small{-0.00} &\small{9.81} &\small{9.81} &\small{78.3}\\
\small{1888} & \small{6} &\small{10.897} &\small{280~s} &\small{0.02} &\small{9.85} &\small{9.83} &\small{93.8}\\
\small{1888} & \small{7} &\small{11.301} &\small{170~s} &\small{-0.00} &\small{9.81} &\small{9.81} &\small{140}\\
\small{1888} & \small{8} &\small{11.211} &\small{1.1~\%} &\small{-0.00} &\small{9.81} &\small{9.81} &\small{131}\\\hline
\end{tabular}
\label{tab:part_time_IMB_zij}
\end{table}

\bibliography{references}

\begin{table}[t]
\centering
\caption{Islands' Power Imbalance Minimization using \eqref{eqn:dc_our}}
\begin{tabular}{|c|c|c|c|c|c|c|c|c|}\hline
 $n$ &  $K$ & $UB$,   	   & $g,~\%$       & $P_{LS}$,   & $P_{\Delta}$,   & $P_{GS}$,    & $p_{ij}^{\Sigma}$,\\
	 &      & p.u.    	   & $T$,~s        & p.u.        & p.u.            & p.u.         & p.u.            \\\hline
\small{89} & \small{2} &\small{7.6364} &\small{19~s} &\small{2.43} &\small{6.24} &\small{3.81} &\small{55.5}\\
\small{89} & \small{3} &\small{7.8258} &\small{52~s} &\small{2.43} &\small{6.24} &\small{3.81} &\small{74.4}\\
\small{89} & \small{4} &\small{13.485} &\small{12~s} &\small{4.88} &\small{11.1} &\small{6.26} &\small{66.6}\\
\small{89} & \small{5} &\small{13.488} &\small{12~s} &\small{4.88} &\small{11.1} &\small{6.26} &\small{66.9}\\\hline
\small{1354} & \small{2} &\small{17.023} &\small{0.4~s} &\small{-0.00} &\small{16.6} &\small{16.6} &\small{22.2}\\
\small{1354} & \small{3} &\small{17.38} &\small{11~s} &\small{-0.00} &\small{16.6} &\small{16.6} &\small{57.9}\\
\small{1354} & \small{4} &\small{17.123} &\small{1.0~s} &\small{-0.00} &\small{16.6} &\small{16.6} &\small{32.3}\\
\small{1354} & \small{5} &\small{17.604} &\small{76~s} &\small{-0.00} &\small{16.6} &\small{16.6} &\small{80.3}\\
\small{1354} & \small{6} &\small{18.089} &\small{180~s} &\small{-0.00} &\small{16.6} &\small{16.6} &\small{129}\\
\small{1354} & \small{7} &\small{21.668} &\small{17~\%} &\small{1.42} &\small{19.5} &\small{18.1} &\small{155}\\
\small{1354} & \small{8} &\small{18.565} &\small{350~s} &\small{-0.00} &\small{16.6} &\small{16.6} &\small{176}\\\hline
\small{1888} & \small{2} &\small{9.9447} &\small{0.5~s} &\small{-0.00} &\small{9.81} &\small{9.81} &\small{3.97}\\
\small{1888} & \small{3} &\small{10.821} &\small{65~s} &\small{-0.00} &\small{9.81} &\small{9.81} &\small{91.7}\\
\small{1888} & \small{4} &\small{10.431} &\small{6.7~s} &\small{0.03} &\small{9.87} &\small{9.84} &\small{45.1}\\
\small{1888} & \small{5} &\small{10.68} &\small{79~s} &\small{-0.00} &\small{9.81} &\small{9.81} &\small{77.5}\\
\small{1888} & \small{6} &\small{10.92} &\small{170~s} &\small{-0.00} &\small{9.81} &\small{9.81} &\small{101}\\
\small{1888} & \small{7} &\small{11.298} &\small{150~s} &\small{-0.00} &\small{9.81} &\small{9.81} &\small{139}\\
\small{1888} & \small{8} &\small{11.208} &\small{283~s} &\small{-0.00} &\small{9.81} &\small{9.81} &\small{131}\\\hline
\end{tabular}
\label{tab:part_time_IMB_our}
\end{table}

\end{document}